\documentstyle[preprint,aps,prd,eqsecnum,twoside,epsfig]{revtex}
\newcommand {\ve}{\varepsilon}
\newcommand {\pr}{\partial}
\newcommand {\prm}{\prime}
\newcommand {\vf}{\varphi}
\newcommand {\cG}{\cal G}
\newcommand {\cD}{\cal D}
\newcommand {\tH}{\tilde H}
\newcommand {\bH}{\bar H}
\newcommand {\bg}{\bar \gamma}
\newcommand {\bk}{\bar \kappa}
\newcommand {\bp}{\bar \psi}

\begin{document}
\baselineskip -16pt
\title{Nonlinear Spinor and Scalar Fields in General Relativity} 
\author{Bijan Saha\\ 
Laboratory of Information Technologies\\ 
Joint Institute for Nuclear Research, Dubna\\ 
141980 Dubna, Moscow region, Russia\\ 
e-mail:  saha@thsun1.jinr.ru, bijan@cv.jinr.ru}

\author{G.N. Shikin \\ 
Department of Theoretical Physics\\ 
Peoples' Friendship University of Russia\\ 
6, Miklukho Maklay Street\\ 
117198 Moscow, Russia}

\maketitle 

\begin{abstract}

We consider a system of nonlinear spinor and scalar fields with
minimal coupling in general relativity. The nonlinearity in the spinor
field Lagrangian is given by an arbitrary function of the invariants 
generated from the bilinear spinor forms $S= \bp \psi$ and 
$P=i \bp \gamma^5 \psi$; the scalar Lagrangian is chosen as an arbitrary
function of the scalar invariant 
${\Upsilon} = {\vf}_{,\alpha}{\vf}^{,\alpha}$,
that becomes linear at ${\Upsilon} \to 0$. The spinor and the scalar 
fields in 
question interact with each other by means of a gravitational field
which is given by a plane-symmetric metric. Exact plane-symmetric
solutions to the gravitational, spinor and scalar field equations
have been obtained. Role of gravitational field in the formation of
the field configurations with limited total energy, spin and charge
has been investigated. Influence of the change of the sign of energy 
density of the spinor and scalar fields on the properties of the
configurations obtained has been examined. It has been established
that under the change of the sign of the scalar field energy density
the system in question can be realized physically iff the scalar 
charge does not exceed some critical value. In case of spinor field no
such restriction on its parameter occurs. In general it has been shown 
that the choice of spinor field nonlinearity can lead to the elimination
of scalar field contribution to the metric functions, but leaving its
contribution to the total energy unaltered.     
\end{abstract}
\vskip 3mm
\noindent
{\bf Key words:} Nonlinear spinor field (NLSF), nonlinear scalar field,
plane-symmetric metric
\vskip 3mm
\noindent
{\bf PACS:} 03.65.P, 04.20.H

\section{Introduction}
Nonlinear phenomena have been one of the most popular topics during 
last years. Nevertheless, it must be admitted that nonlinear classical 
fields have not received general consideration. This is probably due
to the mathematical difficulties which arise because of the 
nonrenormalizability of the Fermi and other nonlinear couplings~\cite{ranada}.

Nonlinear selfcouplings of the spinor fields may arise as a consequence 
of the geometrical structure of the space-time and, more precisely, 
because of the existence of torsion. As soon as 1938, Ivanenko
\cite{ivanenko1,ivanenko2,rodichev}
showed that a relativistic theory imposes in some cases a fourth order
selfcoupling. In 1950 Weyl~\cite{weyl} 
proved that, if the affine and the metric properties of the space-time
are taken as independent, the spinor field obeys either a linear equation
in a space with torsion or a nonlinear one in a Riemannian space. As
the selfaction is of spin-spin type, it allows the assignment of a 
dynamical role to the spin and offers a clue about the origin of the
nonlinearities. This question was further clarified in some important
papers by Utiyama, Kibble and Sciama ~\cite{utiyama,kibble,sciama}
In the simplest scheme the selfaction is of pseudovector
type, but it can be shown that one can also get a scalar coupling
~\cite{soler}. An excellent review of the problem may be found in~\cite{hehl}. 

Nonlinear quantum Dirac fields were used by Heisenberg~\cite{hb1,hb2}
in his ambitious unified theory of elementary particles. They are 
presently the object of renewed interest since the widely known paper
by Gross and Neveu~\cite{gross} 

Nonlinear spinor field (NLSF) in external cosmological
gravitational field was first studied by G.N. Shikin in 1991~\cite{shikin}.  
This study was extended by us for the more general case where 
we consider the nonlinear term as an arbitrary function of all possible 
invariants generated from spinor bilinear forms. In that paper we also
studied the possibility of elimination of initial singularity especially for
the Kasner Universe~\cite{saha1}. For few years we studied 
the behavior of self-consistent NLSF in a B-I Universe~\cite{saha2,saha3} 
both in presence of perfect fluid and without it that was 
followed by the Refs.,~\cite{saha4,saha5,saha6} where we studied
the self-consistent system of interacting spinor and scalar fields.
In a series of paper we also thoroughly studied the interacting
scalar and electromagnetic fields in spherically and cylindrically
space-time~\cite{sizv,sctp,sijtp,sgc,sijmpa}. 
The purpose of the paper is to study the role of nonlinear spinor
and scalar field in the formation of configurations with localized
energy density and limited total energy, spin and charge of the spinor
field.

\section{Fundamental Equations and general solutions}

The Lagrangian of the nonlinear spinor, scalar and gravitational fields
can be written in the form
\begin{equation}
L = \frac{R}{2 \kappa} + L_{sp} + L_{sc}
\label{totlag}
\end{equation}
with 
\begin{equation} 
L_{sp}=\frac{i}{2} 
\biggl[\bp \gamma^{\mu} \nabla_{\mu} \psi- \nabla_{\mu} \bar 
\psi \gamma^{\mu} \psi \biggr] - m\bp \psi + L_N,
\label{splag} 
\end{equation}
and
\begin{equation}
L_{sc} = \Psi ({\Upsilon}), \quad {\Upsilon} = \vf_{,\alpha}\vf^{,\alpha}.
\label{sclag}
\end{equation} 
Here $R$ is the scalar curvature and $\kappa$ is the Einstein's 
gravitational constant. The nonlinear term $L_N$ in spinor Lagrangian 
describes the self-interaction of a spinor field and can be presented 
as some arbitrary functions of invariants generated from the real 
bilinear forms of a spinor field having the form 
$$S\,=\, \bar \psi \psi, \quad                    
P\,=\,i \bar \psi \gamma^5 \psi, \quad
v^\mu\,=\,(\bar \psi \gamma^\mu \psi), \quad
A^\mu\,=\,(\bar \psi \gamma^5 \gamma^\mu \psi), \quad
T^{\mu\nu}\,=\,(\bar \psi \sigma^{\mu\nu} \psi),$$
where $\sigma^{\mu\nu}\,=\,(i/2)[\gamma^\mu\gamma^\nu\,-\,
\gamma^\nu\gamma^\mu]$. 
Invariants, corresponding to the bilinear forms, look like
$$ I = S^2, \quad J = P^2, \quad 
I_v = v_\mu\,v^\mu\,=\,(\bar \psi \gamma^\mu \psi)\,g_{\mu\nu}
(\bar \psi \gamma^\nu \psi),$$ 
$$I_A = A_\mu\,A^\mu\,=\,(\bar \psi \gamma^5 \gamma^\mu \psi)\,g_{\mu\nu}
(\bar \psi \gamma^5 \gamma^\nu \psi), \quad
I_T = T_{\mu\nu}\,T^{\mu\nu}\,=\,(\bar \psi \sigma^{\mu\nu} \psi)\,
g_{\mu\alpha}g_{\nu\beta}(\bar \psi \sigma^{\alpha\beta} \psi).$$ 
According to the Pauli-Fierz theorem,\cite{landau} among the five invariants
only $I$ and $J$ are independent as all other can be expressed by them:
$I_v = - I_A = I + J$ and $I_T = I - J.$ Therefore we choose the nonlinear
term $L_N = F(I, J)$, thus claiming that it describes the nonlinearity
in the most general of its form. 

The  scalar Lagrangian $L_{sc}$ is an arbitrary function of invariant
$\Upsilon = \vf_{,\alpha}\vf^{,\alpha}$, satisfying the condition
\begin{equation}
\lim\limits_{{\Upsilon} \to 0} \Psi ({\Upsilon}) = \frac{1}{2} {\Upsilon} 
+ \cdots
\label{tl}
\end{equation}  
The static plane-symmetric metric we choose in the form
\begin{equation} 
ds^2 = e^{2 \rho} dt^2 - e^{2 \alpha} dx^2 - e^{2 \beta} (dy^2 + dz^2), 
\label{ps}
\end{equation}
where the metric functions $\rho, \alpha, \beta$ depend on the spatial
variable $x$ only and obey the coordinate condition
\begin{equation}
\alpha = 2 \beta + \rho.
\label{cc}
\end{equation}

Variation of (\ref{totlag}) with respect to spinor field $\psi\,(\bp)$
gives nonlinear spinor field equations
\begin{mathletters}
\label{speq}
\begin{eqnarray}
i\gamma^\mu \nabla_\mu \psi - \Phi \psi + i {\cG} \gamma^5 \psi\,&=&\,0, 
\label{speq1} \\
 i \nabla_\mu \bp \gamma^\mu +  \Phi \bp 
- i {\cG}  \bp \gamma^5 \,&=&\,0, \label{speq2}
\end{eqnarray}
\end{mathletters}
with 
$$ \Phi = m - {\cD} = m - 2 S \frac{{\pr F}}{{\pr I}}, \qquad
{\cG} = 2 P \frac{{\pr F}}{{\pr J}},$$
whereas, variation of (\ref{totlag}) with respect to scalar field
yields the following scalar field equation
\begin{equation}
\frac{1}{\sqrt{-g}} \frac{\pr}{\pr x^\nu} \Bigl(\sqrt{-g} g^{\nu\mu}
\frac{d \Psi}{d {\Upsilon}} \vf_{,\mu}\Bigr) = 0.
\label{scfe}
\end{equation}
Varying (\ref{totlag}) with respect to metric tensor
$g_{\mu\nu}$ we obtain the Einstein's field equation 
\begin{equation}
R_{\nu}^{\mu} - \frac{1}{2}\,\delta_{\nu}^{\mu} R = - \kappa T_{\nu}^{\mu}  
\label{ee}
\end{equation} 
which in view of (\ref{ps}) and (\ref{cc}) is written as follows
\begin{mathletters}
\label{BID}
\begin{eqnarray}
G_{0}^{0} = e^{-2\alpha} \bigl(2 \beta^{\prm \prm} -
2 \rho^{\prm}\beta^{\prm} - \beta^{\prm 2}\bigr) = -\kappa T_{0}^{0} 
\label{00}\\ 
G_{1}^{1} = e^{-2\alpha} \bigl(2 \rho^{\prm}\beta^{\prm} 
+ \beta^{\prm 2}\bigr) = -\kappa T_{1}^{1} \label{11}\\ 
G_{2}^{2} = e^{-2\alpha} \bigl(\beta^{\prm \prm} +\rho^{\prm \prm} -
2 \rho^{\prm}\beta^{\prm} - \beta^{\prm 2}\bigr) = -\kappa T_{2}^{2} 
\label{22}\\
G_{3}^{3} = G_{2}^{2}, \quad T_{3}^{3} = T_{2}^{2} \label{23}.
\end{eqnarray}
\end{mathletters}
Here prime denotes differentiation with respect to $x$ and 
$T_{\nu}^{\mu}$ is the energy-momentum tensor of the spinor and
scalar fields 
\begin{equation}
T_{\mu}^{\nu} = T_{{\rm sp}\,\mu}^{\,\,\,\nu} + T_{{\rm sc}\,\mu}^{\,\,\,\nu}.
\label{tem}
\end{equation}
The energy-momentum tensor of the spinor field is 
\begin{equation}
T_{{\rm sp}\,\mu}^{\,\,\,\rho}=\frac{i}{4} g^{\rho\nu} \biggl(\bp \gamma_\mu 
\nabla_\nu \psi + \bp \gamma_\nu \nabla_\mu \psi - \nabla_\mu \bar 
\psi \gamma_\nu \psi - \nabla_\nu \bp \gamma_\mu \psi \biggr) \,-
\delta_{\mu}^{\rho}L_{sp}
\label{temsp}
\end{equation}
where $L_{sp}$ with respect to (\ref{speq}) takes the form
\begin{equation}
L_{sp} = -\frac{1}{2}\Bigl(\bp \frac{\pr F}{\pr \bp} + \frac{\pr F}{\pr \psi}
\psi\Bigr) - F,
\label{lsp}
\end{equation}
and the energy-momentum tensor of the scalar one is
\begin{equation}
T_{{\rm sc}\,\mu}^{\,\,\,\nu}= 2 \frac{d \Psi}{d {\Upsilon}} 
\vf_{,\mu}\vf^{,\nu}
- \delta_{\mu}^{\nu} \Psi,
\quad {\Upsilon} = - (\vf^{\prm})^2 e^{-2 \alpha}, \quad
\vf^{\prm} = \frac{d \vf}{d x}.
\label{temsc}
\end{equation}
In (\ref{speq}) and (\ref{temsp}) 
$\nabla_\mu$ denotes the covariant derivative of spinor, having the 
form~\cite{zhelnorovich,brill}  
\begin{equation} 
\nabla_\mu \psi=\frac{\pr \psi}{\pr x^\mu} -\Gamma_\mu \psi, 
\label{cvd}
\end{equation} 
where $\Gamma_\mu(x)$ are spinor affine connection matrices. 
$\gamma$ matrices in the above equations are connected with 
the flat space-time Dirac matrices $\bg$ in the following way
\begin{equation}
g_{\mu \nu} (x)= e_{\mu}^{a}(x) e_{\nu}^{b}(x) \eta_{ab}, 
\quad \gamma_\mu(x)= e_{\mu}^{a}(x) \bg_a, 
\label{dm}
\end{equation}
where $\eta_{ab}= {\rm diag}(1,-1,-1,-1)$ and $e_{\mu}^{a}$ is a 
set of tetrad 4-vectors. Using (\ref{dm}) we obtain
\begin{equation}
\gamma^0 (x) = e^{-\rho} \bg^0, \quad
\gamma^1 (x) = e^{-\alpha} \bg^1, \quad
\gamma^2 (x) = e^{-\beta} \bg^2, \quad
\gamma^3 (x) = e^{-\beta} \bg^3.
\label{pdm}
\end{equation}
 
From
\begin{equation}
\Gamma_\mu (x)= 
\frac{1}{4}g_{\rho\sigma}(x)\biggl(\partial_\mu e_{\delta}^{b}e_{b}^{\rho} 
- \Gamma_{\mu\delta}^{\rho}\biggr)\gamma^\sigma\gamma^\delta, 
\label{gm}
\end{equation}
one finds
\begin{eqnarray} 
\Gamma_0 = -\frac{1}{2} \bg^0 \bg^1 e^{-2\beta} \rho^{\prm}, \quad 
\Gamma_1 = 0, \quad
\Gamma_2 = \frac{1}{2} \bg^2 \bg^1 e^{-(\rho + \beta)} \beta^{\prm}, \quad 
\Gamma_3 = \frac{1}{2} \bg^3 \bg^1 e^{-(\rho + \beta)} \beta^{\prm}. 
\label{sac}
\end{eqnarray}
Flat space-time matrices $\bg$ we will choose in the form, 
given in~\cite{bogoliubov}:
\begin{eqnarray}
\bg^0&=&\left(\begin{array}{cccc}1&0&0&0\\0&1&0&0\\
0&0&-1&0\\0&0&0&-1\end{array}\right), \quad
\bg^1\,=\,\left(\begin{array}{cccc}0&0&0&1\\0&0&1&0\\
0&-1&0&0\\-1&0&0&0\end{array}\right), \nonumber\\
\bg^2&=&\left(\begin{array}{cccc}0&0&0&-i\\0&0&i&0\\
0&i&0&0\\-i&0&0&0\end{array}\right), \quad
\bg^3\,=\,\left(\begin{array}{cccc}0&0&1&0\\0&0&0&-1\\
-1&0&0&0\\0&1&0&0\end{array}\right).  \nonumber
\end{eqnarray}
Defining $\gamma^5$ as follows,
\begin{eqnarray}
\gamma^5&=&-\frac{i}{4} E_{\mu\nu\sigma\rho}\gamma^\mu\gamma^\nu
\gamma^\sigma\gamma^\rho, \quad E_{\mu\nu\sigma\rho}= \sqrt{-g}
\ve_{\mu\nu\sigma\rho}, \quad \ve_{0123}=1,\nonumber \\
\gamma^5&=&-i\sqrt{-g} \gamma^0 \gamma^1 \gamma^2 \gamma^3 
\,=\,-i\bg^0\bg^1\bg^2\bg^3 =
\bg^5, \nonumber
\end{eqnarray}
we obtain
\begin{eqnarray}
\bg^5&=&\left(\begin{array}{cccc}0&0&-1&0\\0&0&0&-1\\
-1&0&0&0\\0&-1&0&0\end{array}\right).\nonumber
\end{eqnarray}
 
The scalar field equation (\ref{scfe}) has the solution
\begin{equation}
\frac{d \Psi}{d {\Upsilon}} \vf^{\prm} = \vf_{0}, \quad \vf_0 = {\rm const.}
\label{scfs}
\end{equation}
The equality (\ref{scfs}) for a given $\Psi ({\Upsilon})$ is an algebraic
equation for $\vf^{\prm}$ that is to be defined through metric
function $e^{\alpha(x)}$.

We will consider the spinor field to be the function of the spatial
coordinate $x$ only [$\psi=\psi(x)].$ Using (\ref{cvd}), (\ref{pdm}) and
(\ref{sac}) we find
\begin{equation}
\gamma^\mu \Gamma_\mu = -\frac{1}{2} e^{-\alpha} \alpha^{\prm} \bg^1.
\label{Gam}
\end{equation}
Then taking into account (\ref{Gam}) 
we rewrite the spinor field equation (\ref{speq1}) as
\begin{equation} i\bg^1 
\biggl(\frac{\pr}{\pr x} + \frac{\alpha^{\prm}}{2} \biggr) \psi 
+ i e^{\alpha} \Phi \psi +  e^{\alpha}{\cG} \gamma^5 \psi = 0.
\label{spq}
\end{equation} 
Further setting $V(x) = e^{\alpha/2} \psi (x)$ with 
$$V (x) = \left(\begin{array}{c}V_1 (x)\\ V_2 (x)\\ V_3 (x)\\
V_4 (x) \end{array}\right)$$
for the components of spinor field from (\ref{spq}) 
one deduces the following system of equations:  
\begin{mathletters}
\label{V}
\begin{eqnarray} 
V_{4}^{\prm} + i e^{\alpha} \Phi V_1 - e^{\alpha}{\cG} V_3 = 0, \\
V_{3}^{\prm} + i e^{\alpha} \Phi V_2 - e^{\alpha}{\cG} V_4 = 0, \\
V_{2}^{\prm} - i e^{\alpha} \Phi V_3 + e^{\alpha}{\cG} V_1 = 0, \\
V_{1}^{\prm} - i e^{\alpha} \Phi V_4 + e^{\alpha}{\cG} V_2 = 0.
\end{eqnarray} 
\end{mathletters}
As one sees, the equation (\ref{V}) gives following relations 
\begin{equation}
V_{1}^{2} - V_{2}^{2} -V_{3}^{2} +V_{4}^{2} = {\rm const.}
\end{equation}
Using the solutions obtained one can write the components of
spinor current:
\begin{equation}
j^\mu = \bp \gamma^\mu \psi.
\label{spcur}
\end{equation}
Taking into account that $\bp = \psi^{\dagger} \bg^0$, where 
$\psi^{\dagger} = \bigl(\psi_{1}^{*},\,\psi_{2}^{*},\,\psi_{3}^{*},\,
\psi_{4}^{*}\bigr)$ and $\psi_j = e^{-\alpha/2} V_j, \quad j=1,2,3,4$
for the components of spinor current we write
\begin{mathletters}
\label{spincur}
\begin{eqnarray}
j^0 &=& \bigl[V_{1}^{*} V_{1} + V_{2}^{*} V_{2} + V_{3}^{*} V_{3}
+ V_{4}^{*} V_{4}\bigr]e^{-(\alpha + \rho)},\\
j^1 &=& \bigl[V_{1}^{*} V_{4} + V_{2}^{*} V_{3} + V_{3}^{*} V_{2}
+ V_{4}^{*} V_{1}\bigr]e^{-2\alpha},\\
j^2 &=& -i \bigl[V_{1}^{*} V_{4} - V_{2}^{*} V_{3} + V_{3}^{*} V_{2}
- V_{4}^{*} V_{1}\bigr]e^{-(\alpha + \beta)},\\
j^3 &=& \bigl[V_{1}^{*} V_{3} - V_{2}^{*} V_{4} + V_{3}^{*} V_{1}
- V_{4}^{*} V_{2}\bigr]e^{-(\alpha + \beta)}.
\end{eqnarray}
\end{mathletters}
Since we consider the field configuration to be static one, the
spatial components of spinor current vanishes, i.e., 
\begin{equation}
j^1 = 0,\quad  j^2 = 0, \quad j^3 =0. 
\label{suppose}
\end{equation}
This supposition gives additional relation between the constant of 
integration. The component $j^0$ defines the charge
density of spinor field that has the following chronometric-invariant 
form 
\begin{equation}
\varrho = (j_0\cdot j^0)^{1/2}. 
\label{rho}
\end{equation}
The total charge of spinor field is defined as
\begin{equation}
Q = \int\limits_{-\infty}^{\infty} \varrho \sqrt{-^3 g} dx 
\label{charge}
\end{equation}
(in (\ref{charge}) integrations by $y$ and $z$ are performed in the
limit $(0,1)$).

Let us consider the spin tensor~\cite{bogoliubov}
\begin{equation}
S^{\mu\nu,\epsilon} = \frac{1}{4}\bp \bigl\{\gamma^\epsilon
\sigma^{\mu\nu}+\sigma^{\mu\nu}\gamma^\epsilon\bigr\} \psi.
\label{spin}
\end{equation}
We write the components $S^{ik,0}$ $(i,k=1,2,3)$, defining the spatial
density of spin vector explicitly. From (\ref{spin}) we have
\begin{equation}
S^{ij,0} = \frac{1}{4}\bp \bigl\{\gamma^0
\sigma^{ij}+\sigma^{ij}\gamma^0\bigr\} \psi = 
\frac{1}{2}\bp \gamma^0 \sigma^{ij}\psi
\label{spin0}
\end{equation}
that defines the projection of spin vector on $k$ axis. Here $i,\,j,\,k$
takes the value $1,\,2,\,3$ and $i\ne j\ne k$. Thus, for the projection 
of spin vectors on the $X,\,Y$ and $Z$ axis we find
\begin{mathletters}
\label{spinvec}
\begin{eqnarray}
S^{23,0} &=& 
\bigl[V_{1}^{*} V_{2} + V_{2}^{*} V_{1} + V_{3}^{*} V_{4}
+ V_{4}^{*} V_{3}\bigr] e^{-\alpha - 2\beta -\rho},\\
S^{31,0} &=& 
\bigl[V_{1}^{*} V_{2} - V_{2}^{*} V_{1} + V_{3}^{*} V_{4}
- V_{4}^{*} V_{3}\bigr] e^{-2\alpha - \beta -\rho},\\
S^{12,0} &=& 
\bigl[V_{1}^{*} V_{1} - V_{2}^{*} V_{2} + V_{3}^{*} V_{3}
- V_{4}^{*} V_{4}\bigr] e^{-2\alpha - \beta -\rho}.
\end{eqnarray}
\end{mathletters}
The chronometric invariant spin tensor takes the form
\begin{equation}
S_{{\rm ch}}^{ij,0} = \bigl(S_{ij,0} S^{ij,0}\bigr)^{1/2},
\label{chij}
\end{equation} 
and the projection of the spin vector on $k$ axis is defined by
\begin{equation}
S_k = \int\limits_{-\infty}^{\infty} S_{{\rm ch}}^{ij,0} 
\sqrt{-^3 g} dx. 
\label{proj}
\end{equation} 
(In (\ref{proj}), as well as in (\ref{charge}) integrations by $y$ 
and $z$ are performed in the limit $(0,1)$). 

From (\ref{speq}) one can write the equations for 
$ S = \bp \psi, \quad P = i \bp \gamma^5 \psi$ and    
$A = \bp \bg^5 \bg^1 \psi$
\begin{mathletters}
\label{inv}
\begin{eqnarray}
S^{\prm} + \alpha^{\prm} S + 2 e^{\alpha}{\cG}\, A &=& 0, \label{S0}\\
P^{\prm} + \alpha^{\prm} P + 2 e^{\alpha}\Phi\, A &=& 0, \label{P0}\\
A^{\prm} + \alpha^{\prm} A + 2 e^{\alpha}\Phi\, P + 2 e^{\alpha}{\cG} S 
&=& 0. \label{A1} 
\end{eqnarray}
\end{mathletters}
Note that, $A$ in (\ref{inv}) is indeed the pseudo-vector $A^1$. Here for 
simplicity, we use the notation $A$. From (\ref{inv}) immediately follows
\begin{equation}
S^2 + P^2 - A^2 = C_0 e^{-2 \alpha}, \quad C_0 = {\rm const.}
\label{rel}
\end{equation}

Let us now solve the Einstein equations. To do it we first write the 
expression for the components of the energy-momentum tensor explicitly. 
Using the property of flat space-time Dirac matrices and the explicit 
form of covariant derivative $\nabla_\mu$, for the spinor field 
one finds
\begin{equation}
T_{{\rm sp}1}^{\,\,\,1}= m\,S \,-\,F(I,\,J), \quad 
T_{{\rm sp} 0}^{\,\,\,0}=T_{{\rm sp}2}^{\,\,\,2}
=T_{{\rm sp}3}^{\,\,\,3}= {\cD} S + {\cG} P - F(I,\,J). 
\label{temcsp}
\end{equation}
On the other hand, taking into account that the scalar field $\vf$ is
also a function of $x$ only [$\vf = \vf (x)$] for the scalar field one
obtains
\begin{equation}
T_{{\rm sc}1}^{\,\,\,1}= 2 \Upsilon \frac{d \Psi}{d \Upsilon} -
\Psi (\Upsilon), \quad 
T_{{\rm sc} 0}^{\,\,\,0}=T_{{\rm sc}2}^{\,\,\,2}
=T_{{\rm sc}3}^{\,\,\,3}=  - \Psi (\Upsilon). 
\label{temcsc}
\end{equation}

In view of $T_0^0 = T_2^2$,
subtraction of Einstein equations  (\ref{00}) and (\ref{22}) leads  to  
the equation 
\begin{equation}
\beta^{\prm \prm} - \gamma^{\prm \prm} = 0,
\label{bge}
\end{equation} 
with the solution
\begin{equation}
\beta (x) = \gamma (x) + B x,
\label{bgs}
\end{equation}
where $B$ is the integration constant. The second constant has been chosen 
to be trivial, since it acts on the scale of $Y$ and $Z$ axes only. In
account of (\ref{bge}) from (\ref{cc}) one obtains  
\begin{equation} 
\beta^{\prm\prm} = \frac{1}{3} \alpha^{\prm\prm}, \quad 
\gamma^{\prm\prm} = \frac{1}{3} \alpha^{\prm\prm}.
\label{abg}
\end{equation}
Solutions to the equation (\ref{abg}) together with (\ref{cc}) and
(\ref{bgs}) lead to the following expression for $\beta (x)$ and $\gamma (x)$
\begin{equation} 
\beta (x) = \frac{1}{3} \bigl(\alpha (x) + B X\bigr), \quad 
\gamma (x) = \frac{1}{3} \bigl( \alpha (x) - 2 B x\bigr).
\label{abgs}
\end{equation}
Equation (\ref{11}), being the first integral of (\ref{00}) and (\ref{22}),
is a first order differential equation. Inserting $\beta$ and $\gamma$
from (\ref{abgs}) and $T_1^1$ in account of (\ref{tem}), (\ref{temcsp}) and
(\ref{temcsc}) into (\ref{11}) for $\alpha$ one gets
\begin{equation}
\alpha^{\prm 2} - B^2 = -3 \kappa e^{2\alpha} \Bigl[ mS - F(I, J)
+ 2\Upsilon \frac{d \Psi}{d \Upsilon} - \Psi (\Upsilon)\Bigr].
\label{alpha}
\end{equation}
AS one sees from (\ref{inv}) and (\ref{rel}), the invariants are the 
functions of $\alpha$, so is the right hand side of (\ref{alpha}), hence
can be solved in quadrature. In the sections to follow, we analyze 
the equation (\ref{alpha}) in details given the concrete form of 
nonlinear term in spinor Lagrangian.

\section{Analysis of the results} 

In this section we shall analyze the general results obtained in the
previous section for concrete nonlinear term. 

\subsection{Case with linear spinor and scalar fields}

Let us consider the self-consistent system of linear
spinor and massless scalar field equations. By doing so we can
compare the results obtained with those of the self-consistent system 
of nonlinear spinor and scalar field equations, hence clarify the 
role of nonlinearity of the fields in question in the formation
of regular localized solutions such as static solitary wave or 
solitons~\cite{shikin1,shikin2}.

In this case for the scalar field we have $\Psi (\Upsilon) =
\frac{1}{2} \Upsilon$. Inserting this into (\ref{scfs}) we obtain
\begin{equation}
\vf^{\prm} (x) = \vf_0.
\label{lscf}
\end{equation}
From (\ref{temcsc}) in account of (\ref{lscf}) we get
\begin{equation}
- T_{{\rm sc}1}^{\,\,\,1}= T_{{\rm sc} 0}^{\,\,\,0}=
T_{{\rm sc}2}^{\,\,\,2}= T_{{\rm sc}3}^{\,\,\,3}= 
-\frac{1}{2} \Upsilon = \frac{1}{2} \vf_{0}^{2} e^{-2 \alpha}. 
\label{ltemcsc}
\end{equation}
On the other hand for the linear spinor field we have
\begin{equation}
T_{{\rm sp}1}^{\,\,\,1}= m\,S, \quad 
T_{{\rm sp} 0}^{\,\,\,0}=T_{{\rm sp}2}^{\,\,\,2}
=T_{{\rm sp}3}^{\,\,\,3}= 0. 
\label{ltemcsp}
\end{equation}
As one can easily verify, for the linear spinor field the equation
(\ref{S0}) results
\begin{equation}
S = C_0 e^{-\alpha}.
\label{sal}
\end{equation}
Taking this relation into account and the fact that 
$\alpha^{\prm} (x) = -\frac{1}{S}\frac{d S}{d x}$  from
(\ref{alpha}) we write 
\begin{equation}
\int \frac{d S}{\sqrt{(1 + \bk/2) B^2 S^2 -
3 \kappa C_{0}^{2} S}} = x, \quad \bk = 3\kappa \vf_{0}^{2}/B^2,
\label{ISL}
\end{equation}
with the solution
\begin{equation}
S (x) = \frac{M^2}{H^2} {\rm cosh}^2 (\tH x), \quad
M^2 = 3 \kappa C_{0}^{2}, \quad H^2 = B^2(1 + \bk/2), \quad \tH = H/2.
\label{SL}
\end{equation}
Further we define the functions $\psi_j$. Taking into account that 
in this case 
$${\cal F}(S)= m C_0/S\sqrt{H^2 S^2 - M^2 S},$$ 
for $N_{1,2}$ in view of (\ref{SL}) we find 
$$N_{1,2} (x) = \pm (2H/3 \kappa C_0) {\rm tanh} (\tH x) + R_{1,2}.$$
We can then finally write
\begin{eqnarray}
\label{psil}
\psi_{1,2}(x) = i a_{1,2} E(x) {\rm cosh} [f(x) + R_{1,2}],\nonumber \\ \\
\psi_{3,4}(x) = a_{2,1} E(x) {\rm sinh} [f(x) + R_{2,1}],\nonumber
\end{eqnarray}
where $E(x) = \sqrt{3 \kappa m C_0/H^2} {\rm cosh}(\tH x)$ and
$f(x) = (2H/3\kappa C_0) {\rm tanh}(\tH x).$
For the scalar field energy density we find
\begin{equation}
T_{{\rm sc}0}^{\,\,\,0} (x) = \frac{1}{2}\vf_{0}^{2}e^{-2\alpha}
= \frac{M^4 \vf_{0}^{2}}{2C_{0}^{2}H^4}{\rm cosh}^4(\tH x).
\label{scfed}
\end{equation}
It is clear from (\ref{scfed}) that the scalar field energy density
is not localized.

Let us consider the case when the scalar field possesses negative
energy density. Then we have $\Psi (\Upsilon) = -(1/2) \Upsilon$ and
\begin{equation}
- T_{{\rm sc}1}^{\,\,\,1}= T_{{\rm sc} 0}^{\,\,\,0}=
T_{{\rm sc}2}^{\,\,\,2}= T_{{\rm sc}3}^{\,\,\,3}= 
\frac{1}{2} \Upsilon = -\frac{1}{2} \vf_{0}^{2} e^{-2 \alpha}. 
\label{ltemcscn}
\end{equation}
Then for $S$ we get
\begin{equation}
\int \frac{d S}{\sqrt{(1 - \bk/2) B^2 S^2 -
3 \kappa C_{0}^{2} S}} = x.
\label{ISLN}
\end{equation}
As one sees, the field system considered here is physically
realizable iff $1 - \bk/2 > 0$, i.e.,
the scalar charge $|\vf_0| < \sqrt{2/3\kappa} B$. Moreover, 
in the specific case with $B = 0$, independent to the quantity of
scalar charge $\vf_0$, the existence of scalar field with negative
energy density in general relativity is impossible (even in
absence of linear spinor field).

For the total charge $Q$ of the system in this case we have
\begin{equation}
Q = 2 a^2 \int\limits_{-\infty}^{\infty} {\rm cosh}\Bigl[
\frac{4H}{3\kappa C_0} {\rm tanh}(\tH x) + 2 R\Bigr]\Bigl(
\frac{C_0 H^2}{M^2 {\rm cosh}^2 (\tH x)}\Bigr)^{3/2} e^{2 B x/3}\,
dx < \infty.
\end{equation}
It can be shown that, in case of linear spinor and scalar fields 
with minimal coupling both charge and spin of spinor field are
limited. The energy density of the system, in view of (\ref{ltemcsp})
is defined by the contribution of scalar field only:  
\begin{equation}
T_{0}^{0}(x) = T_{{\rm sc} 0}^{\,\,\,0} (x) = \frac{1}{2}
\frac{\vf_{0}^{2} M^4}{C_{0}^{2}H^4} {\rm cosh}^4 (\tH x).
\label{edl}
\end{equation}
From (\ref{edl}) follows that, the energy density of the system
is not localized and the total energy of the system
$E =\int\limits_{-\infty}^{\infty} T_{0}^{0}\sqrt{-^3g} dx$ is
not finite.

\subsection{Nonlinear spinor and linear scalar fields}

{\bf Case I: F = F(I).}
Let us consider the case when the nonlinear term in spinor field
Lagrangian is a function of $I$ ($S$) only, that leads to 
${\cal G}\,=\,0$. From (\ref{inv}) as in case of linear 
spinor field we find $S = C_0 e^{-\alpha (x)}$. Proceeding as
in foregoing subsection, for $S$ from (\ref{alpha}) we write
\begin{equation}
\frac{d S}{d x} = \pm {\cal L}(S), \quad
{\cal L} (S) = \sqrt{B^2 S^2 - 3 \kappa C_0^2 
\Bigl[ mS - F(S) + 2\Upsilon \frac{d \Psi}{d \Upsilon} - 
\Psi (\Upsilon)\Bigr]}
\label{S}
\end{equation}
with the solution
\begin{equation}
\int\frac{d S}{{\cal L}(S)} = \pm (x + x_0).
\label{quadS}
\end{equation}
Given the concrete form of the functions $F(S)$ and $\Psi(\Upsilon)$,
from (\ref{quadS}) yields $S$, hence $\alpha, \beta, \rho$.

Let us now go back to spinor field equations (\ref{V}). Setting
$V_j (x) = U_j (S)$, $j=1,2,3,4$ and taking into account that
in this case ${\cal G} = 0$, for $U_j (S)$ we obtain
\begin{mathletters}
\label{U}
\begin{eqnarray} 
\frac{d U_4}{d S} + i {\cal F}(S) U_1  = 0, \label{U4}\\
\frac{d U_3}{d S} + i {\cal F}(S) U_2  = 0, \label{U3}\\
\frac{d U_2}{d S} - i {\cal F}(S) U_3  = 0, \label{U2}\\
\frac{d U_1}{d S} - i {\cal F}(S) U_4  = 0, \label{U1}
\end{eqnarray} 
\end{mathletters}
with ${\cal F} (S) = \Phi {\cal L}(S) C_0/S.$
Differentiating (\ref{U4}) with respect to $S$ and inserting (\ref{U1})
into it for $U_4$ we find
\begin{equation}
\frac{d^2 U_4}{d S^2} - \frac{1}{{\cal F}} \frac{d {\cal F}}{d S}
\frac{d U_4}{d S} - {\cal F}^2 U_4 = 0
\label{2U4}
\end{equation}
that transforms to
\begin{equation}
\frac{1}{{\cal F}} \frac{d}{d S}\Bigl(\frac{1}{{\cal F}}\frac{d U_4}{d S}
\Bigr) - U_4 = 0,
\label{tr}
\end{equation}
with the first integral
\begin{equation}
\frac{d U_4}{d S} = \pm \sqrt{U_4^2 + C_1}\cdot{\cal F}(S), \quad 
C_1 = {\rm const.}
\label{fi0}
\end{equation}
For $C_1 = a_1^2 > 0$ from (\ref{fi0}) we obtain
\begin{equation}
U_4 (S) = a_1 {\rm sinh} N_1 (S), \quad 
N_1 = \pm \int {\cal F} (S) dS + R_1, \quad R_1 = {\rm const.}
\label{U4g0}
\end{equation}
whereas, for $C_1 = -b_1^2 < 0$ from (\ref{fi0}) we obtain
\begin{equation}
U_4 (S) = a_1 {\rm cosh} N_1 (S)
\label{U4l0}
\end{equation}
Inserting (\ref{U4g0}) and (\ref{U4l0}) into (\ref{U1}) one finds
\begin{equation}
U_1 (S) = i a_1 {\rm cosh} N_1 (S), \quad
U_1 (S) = i b_1 {\rm sinh} N_1 (S).
\label{U10}
\end{equation}
Analogically, for $U_2$ and $U_3$ we obtain
\begin{equation}
U_3 (S) =  a_2 {\rm sinh} N_2 (S), \quad
U_3 (S) =  b_2 {\rm cosh} N_2 (S).
\label{U30}
\end{equation}
and
\begin{equation}
U_2 (S) = i a_2 {\rm cosh} N_2 (S), \quad
U_2 (S) = i b_2 {\rm sinh} N_2 (S).
\label{U20}
\end{equation}
where $N_2 = \pm \int {\cal F} (S) dS + R_2$ and $a_2, b_2$ and $R_2$
are the integration constants. Thus we find the general solutions to   
the spinor field equations (\ref{U}) containing four arbitrary constants.

Using the solutions obtained, from (\ref{spincur}) we find 
the components of spinor current
\begin{mathletters}
\label{spincur1}
\begin{eqnarray}
j^0 &=& \bigl[a_{1}^{2} {\rm cosh} (2 N_1 (S)) + 
a_{2}^{2} {\rm cosh} (2 N_2 (S))\bigr] e^{-(\alpha + \rho)},\\
j^1 &=& 0, \\
j^2 &=& - \bigl[a_{1}^{2} {\rm sinh} (2 N_1 (S)) -
a_{2}^{2} {\rm sinh}(2 N_2 (S))\bigr] e^{-(\alpha + \beta)},\\
j^3 &=& 0.
\end{eqnarray}
\end{mathletters}
The supposition (\ref{suppose}) leads to the following relations between 
the constants: $a_1 = a_2 = a$ and $R_1 = R_2 = R$, since 
$N_1 (S) = N_2 (S) = N(S)$. 
The chronometric-invariant form of the charge density and the total
charge of spinor field are
\begin{equation}
\varrho = 2 a^2 {\rm cosh}(2 N(S)) e^{-\alpha},
\end{equation}
\begin{equation}
Q = 2 a^2 \int\limits_{-\infty}^{\infty} {\rm cosh} (2 N(S)) 
e^{\alpha - \rho} d x.
\label{charge1}
\end{equation}

From (\ref{spin0}) we find
\begin{equation}
S^{12,0} = 0, \quad S^{13,0} = 0,\quad S^{23,0} = a^2 
{\rm cosh} (2 N(S)) e^{-2\alpha}.
\end{equation} 
Thus, the only nontrivial component of the spin tensor is $S^{23,0}$
that defines the projection of spin vector on $X$ axis. From (\ref{chij})
we write the chronometric invariant spin tensor 
\begin{equation}
S_{{\rm ch}}^{23,0} =  a^2 {\rm cosh} (2 N(S)) e^{-\alpha},
\label{ch20}
\end{equation} 
and the projection of the spin vector on $X$ axis 
\begin{equation}
S_1 = a^2 \int\limits_{-\infty}^{\infty}
{\rm cosh} (2 N(S)) e^{\alpha - \rho} d x.
\label{proj1}
\end{equation} 
(in (\ref{proj}), as well as in (\ref{charge}) integrations by $y$ 
and $z$ are performed in the limit $(0,1)$). Note that the integrants
both in (\ref{charge1}) and (\ref{proj1}) coincide. 

Let us now analyze the result obtained choosing the nonlinear term
in the form $F(I) = \lambda S^n = \lambda I^{n/2}$ with $n \ge 2$ and
$\lambda$ is the parameter of nonlinearity.
For $n = 2$ we have Heisenberg-Ivanenko type nonlinear spinor
field equation~\cite{ivanenko}
\begin{equation}
i e^{-\alpha} \bg^1 \bigl(\pr_x +\frac{1}{2} \alpha^{\prm}\bigr) \psi
- m\psi + 2\lambda (\bp \psi )\psi = 0.
\label{hi}
\end{equation}
Setting $F = S^2$ into (\ref{quadS}) we come to the expression 
for $S$ that is similar to that for linear case with 
\begin{equation}
H^2 \to H_1^2 = B^2 + 3 \kappa \lambda C_0 + 3\kappa \vf_{0}^{2}/2.
\label{bH}
\end{equation}

Let us write the functions $\psi_j$ explicitly. In this case we have
$${\cal F}(S)= m (C_0 - 2 \lambda S)/S\sqrt{H_1^2 S^2 - M^2 S},$$ 
and 
$$N_{1,2} (x) = (2 H_1/3 \kappa C_0) {\rm tanh} (\bH_1 x) 
- 2 \lambda C_0 x + R_{1,2}, \quad \bH_1 = H_1/2.$$
We can then finally write
\begin{eqnarray}
\label{psihi}
\psi_{1,2}(x) = i a_{1,2} \frac{\sqrt{3\kappa m C_0}}{H_1}
{\rm cosh} (\bH_1 x) {\rm cosh} N_{1,2} (x), \nonumber \\ \\
\psi_{3,4}(x) = i a_{2,1} \frac{\sqrt{3\kappa m C_0}}{H_1}
{\rm cosh} (\bH_1 x) {\rm cosh} N_{2,1} (x). \nonumber
\end{eqnarray}
Let us consider the energy-density distribution of the field system:
\begin{equation}
T_{0}^{0} = \bigl(\lambda +\frac{1}{2} \frac{\vf_{0}^{2}}{C_{0}^{2}}
\bigr)\frac{M^4}{H_1^4} {\rm cosh}^4 (\bH_1 x). 
\label{edd}
\end{equation}
From (\ref{edd}) follows that, the energy density of the system
is not localized and the total energy of the system
$E =\int\limits_{-\infty}^{\infty} T_{0}^{0}\sqrt{-^3g} dx$ is
not finite. Note that, the energy density of the system can be
trivial, if 
\begin{equation}
\lambda +\frac{1}{2} \frac{\vf_{0}^{2}}{C_{0}^{2}} = 0.
\end{equation}
It is possible, iff the sign of energy density of spinor and 
scalar fields are different. 

Let us write the total charge of the system.
\begin{equation}
Q = 2 a^2 \int\limits_{-\infty}^{\infty} {\rm cosh}\Bigl[
\frac{4 H_1}{3\kappa C_0} {\rm tanh}(\bH_1 x) -
4 \lambda C_0 x + 2 R\Bigr]\Bigl(
\frac{C_0 H_1^2}{M^2 {\rm cosh}^2 (\bH_1 x)}\Bigr)^{3/2} 
e^{2 B x/3}\,
dx.
\label{chhi}
\end{equation}
If $12 \lambda^2 C_{0}^{2} + \lambda C_0 (4 B - \kappa C_0) -
\kappa \vf_{0}^{2}/2 < 0$, the integral (\ref{chhi}) converges,
that means the possibility of existence of finite charge and spin
of the system.

In case of $n > 2$, the energy density of the system in question
is
\begin{equation}
T_{0}^{0} = \lambda (n - 1) S^n + \frac{1}{2}\frac{\vf_{0}^{2}}
{C_{0}^{2}} S^2,
\end{equation}
which shows that the regular solutions with localized energy density
exists iff $S = \bp \psi$ is a continuous and limited function
and $\lim\limits_{ x \to \pm \infty} S(x) \to 0$. The condition, 
when $S$ possesses the properties mentioned above is
\begin{equation}
\int\limits_{}^{} \frac{d S}{\sqrt{(1 + \bk/2) B^2 S^2
- 3 \kappa C_0^2 (m S - \lambda S^n)}} = x.
\label{conS}
\end{equation}
As one sees from (\ref{conS}), for $m \ne 0$ at no value of $x$
$S$ becomes trivial, since as $S \to 0$, the denominator of the 
integrant beginning from some finite value of $S$ becomes imaginary.
It means that for $S(x)$ to be trivial at spatial infinity 
($x \to \infty$), it is necessary to choose massless spinor field
setting $m = 0$ in (\ref{conS}). Note that, in the unified nonlinear 
spinor theory of Heisenberg, the massive term is absent, and 
according 
to Heisenberg, the particle mass should be obtained as a result of 
quantization of spinor prematter~\cite{heisenberg}. It should be 
emphasized that in the nonlinear generalization of classical field 
equations, the massive term does not possess the significance that 
it possesses in the linear one, as it by no means defines 
total energy (or mass) of the nonlinear field system~\cite{schweber}. 
Thus without losing the generality we can consider massless spinor 
field putting $m = 0$. Note that in the sections to follow where
we consider the nonlinear spinor term as $F = P^n$, or $F = (K_\pm)^n$
with $K_\pm = (I \pm J)$, we will study the massless spinor field only.

From (\ref{conS}) for $m = 0$, $\lambda > 0$ and $n > 2$ for $S(x)$ we
obtain
\begin{equation}
S (x) = \Bigl[-H_1/\sqrt{3 \kappa \lambda C_{0}^{2} (\zeta^2 - 1)} 
\Bigr]^{2/(n-2)}, \quad \zeta = {\rm cosh}[(n-2)\bH_1 x]
\end{equation} 
from which follows that $\lim\limits_{x \to 0} |S(x)| \to \infty$. 
It means that
$T_{0}^{0} (x)$ is not bounded at $x = 0$ and the initial system of 
equations does not possess solutions with localized energy density.

If we set in (\ref{conS}) $m = 0$, $\lambda = -\Lambda^ 2 < 0$ and
$n > 2$, then for $S$ we obtain
\begin{equation}
S (x) = \Bigl[H_1/\sqrt{3 \kappa \lambda C_{0}^{2}} \zeta\Bigr]^{2/(n-2)}
\label{negcoup}
\end{equation} 
It is seen from (\ref{negcoup}) that $S(x)$ has maximum at $x = 0$ 
and $\lim\limits_{x \to \pm \infty} S(x) \to 0$. For energy density 
we have
\begin{equation}
T_{0}^{0} = -\Lambda^2 (n - 1) S^n + \frac{1}{2}\frac{\vf_{0}^{2}}
{C_{0}^{2}} S^2,
\label{negend}
\end{equation}
where $S$ is defined by (\ref{negcoup}). In view of $S$ it follows 
that $T_{0}^{0}(x)$ is an alternating function.

Let us find the condition when the total energy of the system is bound
\begin{equation}
E = \int\limits_{-\infty}^{\infty} T_{0}^{0}\sqrt{-^3 g} dx < \infty.
\label{toen}
\end{equation}
For this we write the integrant of (\ref{toen})
\begin{eqnarray}
\varepsilon (x) &=&
\label{integrant}
T_{0}^{0}\sqrt{-^3 g} = \\ 
&=& C_{0}^{5/3}\Bigl[\frac{\vf_{0}^{2}}
{2C_{0}^{2}} - \frac{(n-1)H_1^2 \zeta^2}{3 \kappa \lambda C_{0}^{2}} 
\Bigr]
\Bigl[\frac{H_1^2 \zeta}{3 \kappa \Lambda^2 C_{0}^{2}} 
\Bigr]^{1/3(n-2)}\cdot e^{2Bx/3}.\nonumber
\end{eqnarray}
From (\ref{integrant}) follows that 
$\lim\limits_{x \to -\infty} \varepsilon (x) \to 0$ for any value of the
parameters, while
$\lim\limits_{x \to +\infty} \varepsilon (x) \to 0$ iff
$H > 2 B$ or $\kappa \vf_{0}^{2}> 2 B^2$. Note that in this case 
the contribution of scalar field to the total energy in positive
and finite:
\begin{equation}
T_{{\rm sc}0}^{\,\,\,0} = \frac{\vf_{0}^{2}}{2 C_{0}^{2}} S^2,\quad
E_{\rm sc} = \int\limits_{-\infty}^{\infty} 
T_{{\rm sc}0}^{\,\,\,0}\sqrt{-^3 g} dx < \infty.
\end{equation}
Note that in the case considered the scalar field is linear and
massless. As far as in absence of spinor field energy density
of the linear scalar field is not localized and the total energy 
in not finite, in the case considered the properties of the
field configurations are defined by those of nonlinear spinor field.
The contribution of nonlinear spinor field to the total energy is
negative. Moreover, it remains finite even in absence of scalar
field for $n > 2$ ~\cite{shikin3}.

The components of spinor field in this case have the form
\begin{eqnarray}
\label{psihi2}
\psi_{1,2}(x) = i a_{1,2} E(x) {\rm cosh} N_{1,2} (x), \nonumber\\ \\
\psi_{3,4}(x) = a_{2,1} E(x) {\rm sinh} N_{2,1}(x), \nonumber
\end{eqnarray}
where 
$$E(x) = (1/\sqrt{C_0})\Bigl[H_1/
\sqrt{3 \kappa \Lambda^2 C_0^2} \zeta\Bigr]^{1/(n-2)}$$ 
and
$$N_{1,2} (x) = -\frac{2n H_1\sqrt{\zeta^2 -1}}{3 \kappa C_0 (n-2)\zeta}  
+ R_{1,2}.$$

For the solutions obtained we write the chronometric-invariant
charge density of the spinor field $\varrho$:
\begin{eqnarray}
\varrho (x) = \frac{2 a^2}{C_0}{\rm cosh} \Bigl\{ -
\frac{4 n H_1 \sqrt{\zeta^2 -1}
}{3 \kappa C_0 (n - 2)\zeta}  + 2 R\Bigr\}
\Bigl\{\frac{H_1^2 }{3 \kappa \Lambda^2 C_{0}^{2}\zeta^2}
\Bigr\}^{1/(n-2)}.
\label{chihi}
\end{eqnarray}
As one sees from (\ref{chihi}), the charge density is localized, since
$\lim\limits_{x \to \pm \infty} \varrho (x) \to 0$. Nevertheless,
the charge density of the spinor field, coming to unit invariant
volume $\varrho\sqrt{-^3 g}$, is not localized:
\begin{equation}
\varrho\sqrt{-^3 g} = 2 a^2 {\rm cosh} [2 N(x)]e^{\alpha - \gamma}
= 2 a^2 {\rm cosh} [2 N(x)] (C_0/S)^{2/3} e^{2Bx/3}.
\end{equation}
It leads to the fact that the total charge of the spinor field is not
bounded as well. As far as the expression for chronometric-invariant
tensor of spin (\ref{ch20}) coincides with that of $\varrho (x)/2$,
the conclusions made for $\varrho (x)$ and $Q$ will be valid for
the spin tensor $S_{\rm ch}^{23,0}$ and projection of spin vector
on $X$ axis $S_1$, i.e., $S_{\rm ch}^{23,0}$ is localized and
$S_1$ is unlimited. 

The solution obtained describes the configuration of nonlinear spinor
and linear scalar fields with localized energy density but with the
metric that is singular at spatial infinity, as in this case
\begin{equation}
e^{2\alpha} = (C_0/S)^2 = C_{0}^{2}\Bigl\{\frac{3 \kappa \Lambda
C_{0}^{2}\zeta}{H_1^2}\Bigr\}^{2/(n-2)} 
\Bigl|_{x \to \pm \infty} \to \infty
\end{equation}

Let us consider the massless spinor field with 
\begin{equation}
F = - \Lambda^2 S^{-\nu}, \quad \nu = {\rm constant} > 0.
\label{nu}
\end{equation}
In this case the energy density of the system of nonlinear spinor and 
linear scalar fields with minimal coupling takes the form
\begin{equation}
T_{0}^{0} = \Lambda^2 (\nu + 1) S^{-\nu} + \frac{\vf_0^2}{2 C_0^2} S^2
\end{equation}
For $S$ in this case we get
\begin{equation}
\int\limits_{}^{} \frac{d S}{\sqrt{(1 + \bk/2) B^2 S^2
- 3 \kappa C_0^2 \Lambda^2 S^{-\nu}}} = x
\end{equation}
with the solution
\begin{equation}
S (x) = \Bigl[\frac{3 \kappa \Lambda^2 C_{0}^{2}}{H_1^2} \zeta_1^2  
\Bigr]^{1/(\nu + 2)}, \quad
\zeta_1 = {\rm cosh} [(\nu + 2) \bH_1 x].
\end{equation} 
For energy density in this case we have
\begin{eqnarray}
T_{0}^{0} (x) = \Lambda^2 (\nu + 1) 
\Bigl[\frac{H_1^2}{3 \kappa C_{0}^{2}\Lambda^2 \zeta_1^2 
}\Bigr]^{\nu/(\nu+2)} +\frac{\vf_{0}^{2}}{2 C_{0}^{2}}
\Bigl[\frac{3 \kappa C_{0}^{2}\Lambda^2  \zeta_1^2
}{H_1^2}\Bigr]^{2/(\nu+2)}.
\label{ednu}
\end{eqnarray}
It follows from (\ref{ednu}) that the contribution of the spinor
field in the energy density is localized while for the scalar field
it is not the case.

The energy density distribution of the field system, coming to
unit invariant volume is
\begin{eqnarray}
\varepsilon (x) &=&
T_{0}^{0}\sqrt{-^3 g} = \Bigl[\Lambda^2 (\nu + 1) S^{-\nu}
+ \frac{\vf_{0}^{2}}{2C_{0}^{2}} S^2 \Bigr] e^{2\alpha - \gamma}
\nonumber\\ 
& = & \Bigl\{\frac{H_1^2 (\nu + 1)}{3\kappa \zeta_1^2 
} +\frac{\vf_{0}^{2}}{2}\Bigr\}\Bigl\{\frac{H_1^2}
{3\kappa C_{0}^{2} \Lambda^2 \zeta_1^2}\Bigr\}^{1/3(\nu+2)} e^{2Bx/3}.
\label{integrant1}
\end{eqnarray}
As one sees from (\ref{integrant1}) $\varepsilon (x)$ is a localized 
function, i.e., $\lim\limits_{x \to \pm \infty} \varepsilon (x) \to 0$,
if $H > 2 B$ or $\kappa \vf_{0}^{2} > 2 B^2$. In this case the total
energy is also finite.

The components of spinor field in this case have the form
\begin{eqnarray}
\label{psihi3}
\psi_{1,2}(x) = i a_{1,2} E(x) {\rm cosh} N_{1,2} (x), \nonumber \\ \\
\psi_{3,4}(x) = a_{2,1} E(x) {\rm sinh} N_{2,1}(x), \nonumber
\end{eqnarray}
where 
$$E(x) = (1/\sqrt{C_0})\Bigl[\frac{
\sqrt{3 \kappa \Lambda^2 C_0^2}}{H_1^2}\zeta_1\Bigr]^{1/(\nu+2)}$$ 
and
$$N_{1,2} (x) = -\frac{2H\nu\sqrt{\zeta_1^2 -1}}
{3 \kappa C_0 (\nu + 2)\zeta_1} + R_{1,2}.$$

The chronometric-invariant charge density of the spinor field coming
to unit invariant volume with $a_1 = a_2 = a$ and $N_1 = N_2$ reads
\begin{eqnarray}
\label{vr0}
\varrho\sqrt{-^3 g} &=& 2 a^2 {\rm cosh} [2 N(x)]e^{\alpha - \gamma}= 
\\
&=& 2 a^2 (C_0)^{2/3} {\rm cosh} \Bigl\{2R - 
\frac{4 H_1 \nu \sqrt{\zeta_1^2 -1}}
{3 \kappa C_0 (\nu + 2)\zeta_1} \Bigr\}
\Bigl\{\frac{H_1^2}{3 \kappa C_{0}^{2}\Lambda^2 \zeta_1^2
}\Bigr\}^{2/3(\nu+2)} e^{2Bx/3}.\nonumber
\end{eqnarray}
It follows from (\ref{vr0}) that $\varrho\sqrt{-^3 g}$ is a localized
function and the total charge $Q$ is finite. The spin of spinor field
is limited as well.

{\bf Case II: F = F(J).} Here we consider the massless spinor field
with the nonlinearity $F = F(J)$. In this case from (\ref{P0})
immediately follows
\begin{equation}
P = D_0 e^{-\alpha (x)}, \quad D_0 = {\rm const.}
\label{pal}
\end{equation}
From (\ref{V}) we now have
\begin{mathletters}
\label{VP}
\begin{eqnarray} 
V_{4}^{\prm} - e^{\alpha}{\cG} V_3 = 0, \\
V_{3}^{\prm} - e^{\alpha}{\cG} V_4 = 0, \\
V_{2}^{\prm} + e^{\alpha}{\cG} V_1 = 0, \\
V_{1}^{\prm} + e^{\alpha}{\cG} V_2 = 0,
\end{eqnarray} 
\end{mathletters}
with the solutions
\begin{mathletters}
\label{VCP}
\begin{eqnarray} 
V_1 &=& C_1 {\rm sinh}  [-{\cal A} + C_2]\\
V_2 &=& C_1 {\rm cosh}  [-{\cal A} + C_2]\\
V_3 &=& C_3 {\rm sinh}  [{\cal A} + C_4]\\
V_4 &=& C_3 {\rm cosh}  [{\cal A} + C_4]
\end{eqnarray} 
\end{mathletters}
with $C_1, C_2, C_3$ and $C_3$ being the constant of integration
and ${\cal A} = \int e^\alpha {\cG} dx.$

Using the solutions obtained, from (\ref{spincur}) we now find 
the components of spinor current 
\begin{mathletters}
\label{spincur2}
\begin{eqnarray}
j^0 &=& \bigl[C_1^2 {\rm cosh} [2 (-{\cal A} + C_2)]
+ C_3^2 {\rm cosh} [2 ({\cal A} + C_4)] \bigr] e^{-(\alpha + \rho)},\\
j^1 &=& \bigl[2 C_1 C_3 {\rm sinh}(C_2 + C_4)\bigr] e^{-2\alpha}, \\
j^2 &=& 0,\\
j^3 &=& -\bigl[2 C_1 C_3 {\rm cosh}[2 {\cal A} -
C_2 + C_4)\bigr] e^{-(\alpha + \beta)}.
\end{eqnarray}
\end{mathletters}
The supposition (\ref{suppose}) that the spatial components of the
spinor current are trivial leads at least one of the  
constants $(C_1,\,C_3)$ to be zero. Let us set $C_1 = 0.$ 
The chronometric-invariant form of the charge density and the total
charge of spinor field are
\begin{equation}
\varrho = C_{3}^{2} {\rm cosh} [2({\cal A} + C_4)] e^{-\alpha}, 
\end{equation}
\begin{equation}
Q = C_3^2 \int\limits_{-\infty}^{\infty} 
{\rm cosh} [2({\cal A} + C_4)]
e^{\alpha - \rho} d x.
\end{equation}

From (\ref{spin0}) we find
\begin{equation}
S^{12,0} = -C_3^2 e^{-(2 \alpha + \beta + \rho)}, \quad 
S^{31,0} = 0,\quad 
S^{23,0} = C_3^2 {\rm sinh} [2({\cal A} + C_4)] e^{-2\alpha}.
\end{equation} 
Thus, in this case we have two
nontrivial components of the spin tensor $S^{23,0}$ and $S^{12,0}$.
those define the projections of spin vector on $X$ and $Z$ axis,
respectively. From (\ref{chij})
we write the chronometric invariant spin tensor 
\begin{mathletters}
\begin{eqnarray}
S_{{\rm ch}}^{23,0} &=&  C_3^2 
{\rm sinh} [2({\cal A} + C_4)\bigr] e^{-\alpha},\\
S_{{\rm ch}}^{23,0} &=&  C_3^2 e^{-\alpha}
\end{eqnarray}
\end{mathletters}
and the projections of the spin vector on $X$ and $Z$ axes are
\begin{mathletters}
\begin{eqnarray}
S_1 &=& C_3^2 \int\limits_{-\infty}^{\infty} 
{\rm sinh} [2({\cal A} + C_4)]
e^{\alpha - \rho} d x,\\
S_3 &=& C_3^2 \int\limits_{-\infty}^{\infty} e^{\alpha - \rho} d x.
\end{eqnarray}
\end{mathletters} 

Note that the equation for $\alpha$, therefore for $P$ will be the 
same as in previous case (i.e., for $S$ with $m = 0$) with all the
conclusions made there. So we will not proceed further with this. 
We also note that for $F = K_\pm$ with $K_\pm = I \pm J$ for 
massless spinor field we obtain $K_\pm = K_0 e^{-2 \alpha}$ and
the conclusions made above will be remain valid. 

\subsection{Nonlinear scalar field in absence of spinor one} 

Let us consider the system of gravitational and 
nonlinear scalar fields. As a nonlinear scalar field equation we 
choose Born-Infeld one, given by the Lagrangian~\cite{shikin2}
\begin{equation}
\Psi (\Upsilon) = -\frac{1}{\sigma} (1 - \sqrt{1 + \sigma \Upsilon}),
\label{born}
\end{equation}
with $\Upsilon = \vf_\alpha \vf^\alpha$ and $\sigma$ is the
parameter of nonlinearity. From (\ref{born}) we also have
\begin{equation}
\lim\limits_{\sigma \to 0} \Psi (\Upsilon) = \frac{1}{2} \Upsilon \cdots
\label{lapp}
\end{equation}
Inserting (\ref{born}) into (\ref{scfs}) for the scalar field we obtain
the equation
\begin{equation}
\vf^{\prm} (x) = \frac{\vf_0}{\sqrt{1 + \sigma \vf_0^2 e^{-2 \alpha (x)}}},
\label{bisc}
\end{equation}
that gives
\begin{equation}
\Upsilon = -(\vf^{\prm})^2 e^{-2\alpha} = 
-\frac{\vf_0^2 e^{-2 \alpha (x)}}{1 + \sigma \vf_0^2 e^{-2 \alpha (x)}}.
\label{Ups}
\end{equation}
From(\ref{bisc}) follows that $\vf^{\prm}\bigl|_{\sigma = 0} = \vf_0$.

For the case considered in this section we have
\begin{equation}
T_{{\rm sc}0}^{\,\,\,0} =  T_{{\rm sc}2}^{\,\,\,2} =
T_{{\rm sc}3}^{\,\,\,3} = - \Psi (\Upsilon) = \frac{1}{\sigma}
\bigl(1 - 1/\sqrt{1 + \sigma \vf_0^2 e^{-2 \alpha (x)}}\bigr),
\label{023}
\end{equation}
and
\begin{equation}
T_{{\rm sc}1}^{\,\,\,1} = 2 \Upsilon \frac{d \Psi}{d \Upsilon} - \Psi
= \frac{1}{\sigma}
\bigl(1 - \sqrt{1 + \sigma \vf_0^2 e^{-2 \alpha (x)}}\bigr).
\label{11sc}
\end{equation}
Putting (\ref{11sc}) into (\ref{alpha}), in account of $m = 0$ and
$F(I,J) \equiv 0$ for $\alpha$ we find
\begin{equation}
\alpha^{\prm} = \pm \sqrt{B^2 - \frac{3\kappa}{\sigma}e^{2\alpha} 
\bigl(1 - \sqrt{1 + \sigma \vf_0^2 e^{-2 \alpha (x)}}\bigr)}. 
\label{alsc}
\end{equation}
From(\ref{alsc}) one finds
\begin{eqnarray}
& &\int \frac{d \alpha}{
\sqrt{B^2 - \frac{3\kappa}{\sigma}e^{2\alpha} 
\bigl(1 - \sqrt{1 + \sigma \vf_0^2 e^{-2 \alpha (x)}}\bigr)}} =
- \frac{2}{B} {\rm ln}\bigl|\xi + 
\sqrt{\bk+ \xi^2}\bigr| + \nonumber\\
&+& \frac{1}{B \sqrt{1 + \bk/2}} \Biggl[{\rm ln}\Bigl|
\sqrt{2} B \sqrt{\bk+ \xi^2} + \sqrt{2} B
\sqrt{1 + \bk \xi/2}\Bigr| -
{\rm ln} \Bigl|\sqrt{3 \kappa \vf_0^2 (\xi^2 - 2)}\Bigr|\Biggr] = x,
\label{alint}
\end{eqnarray}
with $\xi^2 = 1 + \sqrt{1 + \sigma \vf_0^2 e^{-2 \alpha (x)}}.$
As one sees from (\ref{alint})
\begin{eqnarray}
e^{2\alpha(x)}\Bigl|_{x \to + \infty} &\approx& 
\frac{\sigma \vf_{0}^{2}}{2} e^{2 \sqrt{1 + \bk/2} Bx}
\to \infty, \label{inf} \\
e^{2\alpha(x)}\Bigl|_{x \to - \infty} &\approx& 
\frac{\sigma \vf_{0}^{2}}{2} e^{2 B x} \to 0.
\label{zer}
\end{eqnarray}
Let us study the energy density distribution of nonlinear scalar
field. From (\ref{023}) we find
\begin{equation}
T_{{\rm sc} 0}^{\,\,\,0} (x)\Bigl|_{x = - \infty} = \frac{1}{\sigma},\quad
T_{{\rm sc} 0}^{\,\,\,0} (x)\Bigl|_{x =  \infty} = 0,
\label{0023}
\end{equation}
which shows that the energy density of the scalar field is not
localized. Nevertheless, the energy density on unit invariant volume
is localized if $\kappa \vf_{0}^{2} > 2 B^2$:
\begin{equation}
\varepsilon (x) = T_{{\rm sc} 0}^{\,\,\,0} \sqrt{-^3g} =
\frac{1}{\sigma}\Bigl(1 - \frac{1}{1 + \sigma \vf_{0}^{2}e^{-2\alpha}}
\Bigr)e^{5\alpha/3 + 2 B x/3}\Biggl|_{x \to \pm \infty} \to 0.
\end{equation}

In this case the total energy of the scalar field is also bound.
From (\ref{Ups}) in account of (\ref{inf}) and (\ref{zer}) we also have
\begin{equation}
\Upsilon (x) \Bigl|_{x = - \infty} = -\frac{1}{\sigma}, \quad
\Upsilon (x) \Bigl|_{x =  +\infty} = 0,
\label{Upb}
\end{equation}
showing that $\Upsilon (x)$ is kink-like.

\subsection{Nonlinear spinor and nonlinear scalar field}
 
Finally we consider the self-consistent system of 
nonlinear spinor and scalar fields. We choose the self-action of
the spinor field as $F = \lambda S^n$, $n > 2$, where as the scalar field 
is taken in the form (\ref{born}). Using the line of reasoning 
mentioned earlier, we conclude that the spinor field considered here
should be massless. Taking into account that 
$e^{-2\alpha} = S^2/C_{0}^{2}$ for $S$ we write
\begin{equation}
\int\frac{dS}{\sqrt{B^2 + 3\kappa C_{0}^{2}\bigl[\lambda S^n +
(\sqrt{1 + \sigma \vf_{0}^{2} S^2/C_{0}^{2}} - 1)/\sigma\bigr]}} = x.
\label{nls}
\end{equation} 
From (\ref{nls}) one estimates
\begin{equation}
S(x)\Bigl|_{x \to 0} \sim \frac{1}{x^{2/(n-2)}} \to \infty.
\label{Sin}
\end{equation}
On the other hand for the energy density we have
\begin{equation}
T_{0}^{0} = \lambda (n - 1)S^n + \frac{1}{\sigma}\Bigl(1 -
1/\sqrt{1 + \sigma \vf_{0}^{2} S^2/C_{0}^{2}}\Bigr)
\label{edL}
\end{equation}
that states that for $T_{0}^{0}$ to be localized $S$ should be
localized too and $\lim\limits_{x \to \pm \infty} S(x) \to 0$. 
Hence from (\ref{Sin}) we conclude that $S(x)$ is singular and
energy density in unlimited at $x = 0$.
 
For $\lambda = -\Lambda^2$ and $n > 2$ we have
\begin{equation}
\int\frac{dS}{\sqrt{B^2 + 3\kappa C_{0}^{2}\bigl[-\Lambda^2 S^n +
(\sqrt{1 + \sigma \vf_{0}^{2} S^2/C_{0}^{2}} - 1)/\sigma\bigr]}} = x.
\label{SLam}
\end{equation}
In this case $S(x)$ is finite and its maximum value is defined from
\begin{equation}
S^n (x) =\frac{1}{3\kappa C_{0}^{2}\Lambda^2}\bigl[
B^2 S^2 + 3\kappa C_{0}^{2}
(\sqrt{1 + \sigma \vf_{0}^{2} S^2/C_{0}^{2}} - 1)/\sigma\bigr].
\label{max}
\end{equation}
Noticing that at spatial infinity effects of nonlinearity vanish,
from (\ref{SLam}) we find
\begin{equation}
S(x)\Bigl|_{x \to - \infty} \sim e^{Hx} \to 0, \quad
S(x)\Bigl|_{x \to + \infty} \sim e^{-Hx} \to 0,
\label{ass}
\end{equation}
with $H = \sqrt{B^2 + 3\kappa \vf_{0}^{2}/2} = B\sqrt{1 + \bk/2}$.
In this case the energy density $T_{0}^{0}$ defined by (\ref{edL})
is localized and the total energy of the system in bound. Nevertheless,
spin and charge of the system unlimited.

Let us go back to the general case. For $F = F(S)$ we now have
\begin{equation}
T_{1}^{1} = m S - F(S) + 2\Upsilon \frac{d \Psi}{d \Upsilon} - \Psi.
\label{gc11}
\end{equation} 
It follows that for the arbitrary choice of $\Psi(\Upsilon)$, obeying
(\ref{tl}), we can always choose nonlinear spinor term that will 
eliminate the scalar field contribution in $T_{1}^{1}$, i.e., by virtue 
of total freedom we have here to choose $F(S)$, we can write
\begin{equation}
F(S) = F_1(S) + F_2(S), \quad 
F_2(S) = 2 \Upsilon \frac{d\Psi}{d \Upsilon} - \Psi,
\label{arbit}
\end{equation}
since $\Upsilon = \Upsilon (S^2)$. To prove this we go back to
(\ref{scfs} that gives
\begin{equation}
\Upsilon \Bigl(\frac{d\Psi}{d \Upsilon}\Bigr)^2 = 
-\frac{\vf_{0}^{2} S^2}{C_{0}^{2}}.
\label{alp}
\end{equation}
Since $\Psi$ is the function of $\Upsilon$ only, (\ref{alp}) 
comprises an algebraic equation for defining $\Upsilon$ as a function
of $S^2$. For (\ref{arbit}) takes place, we find
\begin{equation}
(\alpha^{\prm})^2 - B^2 = - \frac{3\kappa C_0^2}{S^2}\bigl[mS - 
F_1(S)\bigr].
\label{alar}
\end{equation}
As we see, the scalar field has no effect on space-time, but it
contributes to energy density and total energy of the system as
in this case
\begin{equation}
T_{0}^{0} = S F_{1}^{\prm}(S) - F_1(S) + S \frac{d}{d\Upsilon}
\bigl(-2 \Upsilon \frac{d\Psi}{d \Upsilon} + \Psi\bigr)
\frac{d \Upsilon}{d S} + 2 \Upsilon \frac{d\Psi}{d \Upsilon} - \Psi.
\end{equation} 
Note that in (\ref{gc11}) with $F(S)$ arbitrary, we cannot choose
$\Psi(\Upsilon)$ such that
\begin{equation}
2 \Upsilon \frac{d\Psi}{d \Upsilon} - \Psi = F(S),
\end{equation}
due to the fact that $\Psi(\Upsilon)$ is not totally arbitrary,
since it has to obey 
\begin{equation}
\lim\limits_{\Upsilon \to 0}\Psi (\Upsilon) \to \frac{1}{2}\Upsilon,
\quad
\lim\limits_{\Upsilon \to 0}
2 \Upsilon \frac{d\Psi}{d \Upsilon} - \Psi = \frac{1}{2}\Upsilon
= \frac{\vf_{0}^{2}}{2 C_{0}^{2}} S^2
\end{equation}
whereas at $S \to 0$, $F(S)$ behaves arbitrarily.

\section{Conclusion}
The system of nonlinear spinor and nonlinear scalar fields with
minimal coupling has been thoroughly studied within the scope
of general relativity given by a plane-symmetric space-time.
Contrary to the scalar field, the spinor field nonliearity 
has direct effect on space-time. Energy density and the total 
energy of the linear spinor and scalar field system are not
bounded and the system does not possess real physical infinity,
hence the configuration is not observable for an infinitely 
remote observer, since in this case 
\begin{equation}
R = \int\limits_{-\infty}^{\infty} \sqrt{g_{11}} dx =
\int\limits_{-\infty}^{\infty} e^\alpha dx = \frac{4 C_0 H}{M^2}
< \infty.
\end{equation}
But introduction of nonliear spinor term into the system 
eliminates these shortcomings and we have the configuration
with finite energy density and limited total energy which is
also observable as in this case the system possesses real
physical infinity. Thus we see, spinor field nonlinearity 
is crucial for the regular solutions with localized energy density. 
We also conclude that the properties of nonlinear spinor and scalar 
field system with minimal coupling are defined by that part of 
gravitational field which is generated by nonlinear spinor one.


\begin{thebibliography}{99}
\bibitem{ranada}
A.F. Ranada, in A.O. Barut (ed.) ``Quantum theory, groups, fields 
and particles'', 271 - 291. Reidel (1983)

\bibitem{ivanenko1} D. Ivanenko, Soviet Physics, {\bf 13}, 141-150 (1938)

\bibitem{ivanenko2} D. Ivanenko, in Einstein Centenay Jubilee Volume 

\bibitem{rodichev} V. Rodichev, Soviet Physics JETP, {\bf 13}, 1029-1033 
(1961)

\bibitem{weyl} H. Weyl, Physical Review, {\bf 77}, 699-701 (1950)

\bibitem{utiyama} R. Utiyama, Physical Review, {\bf 101}, 1596-1607 (1956)

\bibitem{kibble} T.W.B. Kibble, Journal of Mathematical Physics, 
{\bf 2}, 212-221 (1960)

\bibitem{sciama} D.W. Sciama, Festschrift for Infeld, Pergamon Press, 
1960, 415-439. 

\bibitem{soler} A.F. Ranada and M. Soler, Journal of Mathematical Physics, 
{\bf 13}, 671-676 (1972). 

\bibitem{hehl} F.W. Hehl, P. von der Heyde and G.D. Kerlick, 
Review of Modern Physics, {\bf 43}, 393-416 (1976).

\bibitem{hb1} W. Heisenberg, Physica {\bf 19}, 897-908 (1953) 

\bibitem{hb2} W. Heisenberg, Review of Modern Physics, {\bf 29}, 
269-278 (1957)].  

\bibitem{gross} D.J. Gross and A. Neveu, Physical Review D, {\bf 10}, 
3235-3253 (1974)

\bibitem{shikin} G.N. Shikin, Preprint IPBRAE, Academy of Science
USSR {\bf 19}, 21 p. (1991).

\bibitem{saha1} Yu.P. Rybakov,\, B. Saha and G.N. Shikin, 
PFU Reports, Physics, {\bf 2}, (2), 61 (1994).

\bibitem{saha2} Yu.P. Rybakov,\, B. Saha and G.N. Shikin,
Communications in Theoretical Physics {\bf 3}, 199 (1994).

\bibitem{saha3} B. Saha and G.N. Shikin,
Journal of Mathematical Physics {\bf 38}, 5305 (1997).

\bibitem{saha4} R. Alvarado,\, Yu.P. Rybakov,\, B. Saha and G.N. 
Shikin, JINR Preprint {\bf E2-95-16}, 11 p. (1995), 
Communications in Theoretical Physics {\bf 4}, (2), 247 (1995), 
gr-qc/9603035.

\bibitem{saha5} R. Alvarado,\, Yu.P. Rybakov,\, B. Saha and 
G.N. Shikin, Izvestia VUZob, Fizika {\bf 38},(7) 53 (1995).

\bibitem{saha6} B. Saha, and G.N. Shikin, General 
Relativity and Gravitation {\bf 29}, 
1099 (1997).

\bibitem{sizv} Yu.P. Rybakov, B. Saha, and G.N. Shikin,  
Izvestia VUZov, Fizika {\bf 35} (10), 112 (1992). 

\bibitem{sctp} Yu.P. Rybakov, B. Saha, and G.N. Shikin,  
Communications in Theoretical Physics {\bf 3} (1), 67 (1994).

\bibitem{sijtp} Yu.P. Rybakov, B. Saha, and G.N. Shikin,
International Journal of Theoretical Physics {\bf 36} (6), 1475 (1997). 

\bibitem{sgc} Yu.P. Rybakov, B. Saha, and G.N. Shikin,
Gravitation $\&$ Cosmology {\bf 4} (2(14)), 114 (1998).

\bibitem{sijmpa} Bijan Saha,
International Journal of Modern Physics A {\bf 15}(10), 1481 (2000).

\bibitem{einstein} A. Einstein, Sitz. Ber. Preuss. Acad. Wiss.
(1917) and (1919) [English translation: {\it The Principle of
Relativity} (Methuen, 1923, reprinted by Dover, New York, 1924), 
pp. 177 and 191. 

\bibitem{landau} V.B. Berestetski, E.M. Lifshitz and L.P. Pitaevski,
{\it Quantum Electrodynamics} (Nauka, Moscow, 1989).

\bibitem{zhelnorovich} V.A. Zhelnorovich, {\it Spinor theory and its 
application in physics and mechanics} (Nauka, Moscow, 1982).

\bibitem{brill} D. Brill, and J. Wheeler, Review of Modern Physics 
{\bf 29}, 465 (1957).

\bibitem{bogoliubov} N.N. Bogoliubov, and D.V. Shirkov,  
{\it Introduction to the theory of quantized fields} (Nauka, Moscow, 1976).

\bibitem{kamke} E. Kamke, {\it Differentialgleichungen losungsmethoden
und losungen} (Leipzig, 1957). 

\bibitem{shikin1} G.N. Shikin, {\it Basics of soliton theory in 
general relativity} (URSS Publishers, Moscow, 1995).

\bibitem{shikin2} A. Adomou, and G.N. Shikin, Izvestia VUZov, Fizika,
{\bf 41} N. 7, 69 (1998).

\bibitem{ivanenko} D.D. Ivanenko, in {\it Nonlinear quantum field theory} 
(IL, Moscow, 1959) pp. 5 - 40.

\bibitem{heisenberg} W. Heisenberg, {\it An introduction to
unified field theory of elementary particles} (Mir, Moscow, 1968).

\bibitem{schweber} S.S. Schweber, {\it An introduction to 
relativistic quantum field theory} (New York, NY, Harper $\&$ Row, 1961.)

\bibitem{shikin3} A. Adomou, and G.N. Shikin, Gravitation $\&$ Cosmology, 
{\bf 4} N. 2(14), 107 (1998).
\end{thebibliography}
\end{document}